\documentclass{article}
\usepackage{spconf,amsmath,graphicx}
\usepackage{booktabs}
\usepackage{enumitem}
\usepackage{paralist}
\usepackage{multirow}
\usepackage{float}
\usepackage[colorlinks]{hyperref}
\hypersetup{citecolor=black, linkcolor=black}
\setlist{nosep, leftmargin=14pt}
\usepackage{mwe}
\usepackage[capitalize]{cleveref}
\crefname{section}{Sec.}{Secs.}
\Crefname{section}{Section}{Sections}
\Crefname{table}{Table}{Tables}
\crefname{table}{Tab.}{Tabs.}
\title{Head Motion Degrades Machine Learning Classification of Alzheimer's Disease from Positron Emission Tomography}
\name{El\'{e}onore V. Lieffrig$^1$, Takuya Toyonaga$^2$, Jiazhen Zhang$^1$, John A. Onofrey$^{1,2,3}$}
\address{Departments of $^1$Biomedical Engineering, $^2$Radiology and Biomedical Imaging and $^3$ Urology,\\  Yale University, New Haven, CT, USA}
\begin{document}
\maketitle
\begin{abstract}
Brain positron emission tomography (PET) imaging is broadly used in research and clinical routines to study, diagnose, and stage Alzheimer's disease (AD). 
However, its full potential cannot be fully exploited yet due to the lack of portable motion correction solutions, especially in clinical settings.
Head motion during data acquisition has indeed been shown to degrade image quality and induces tracer uptake quantification error.
In this study, we demonstrate that it also biases machine learning-based AD classification. 
We start by proposing a binary classification algorithm solely based on PET images. 
We find that it reaches a high accuracy in classifying motion corrected images into cognitive normal or AD.
We demonstrate that the classification accuracy substantially decreases when images lack motion correction, thereby limiting the algorithm’s effectiveness and biasing image interpretation. 
We validate these findings in cohorts of 128 $^{11}$C-UCB-J and 173 $^{18}$F-FDG scans, two tracers highly relevant to the study of AD.
Classification accuracies decreased by 10\% and 5\% on 20 $^{18}$F-FDG and 20 $^{11}$C-UCB-J testing cases, respectively.
Our findings underscore the critical need for efficient motion correction methods to make the most of the diagnostic capabilities of PET-based machine learning.
\begin{figure}[t!]
    \centering
    \includegraphics[width=\linewidth]{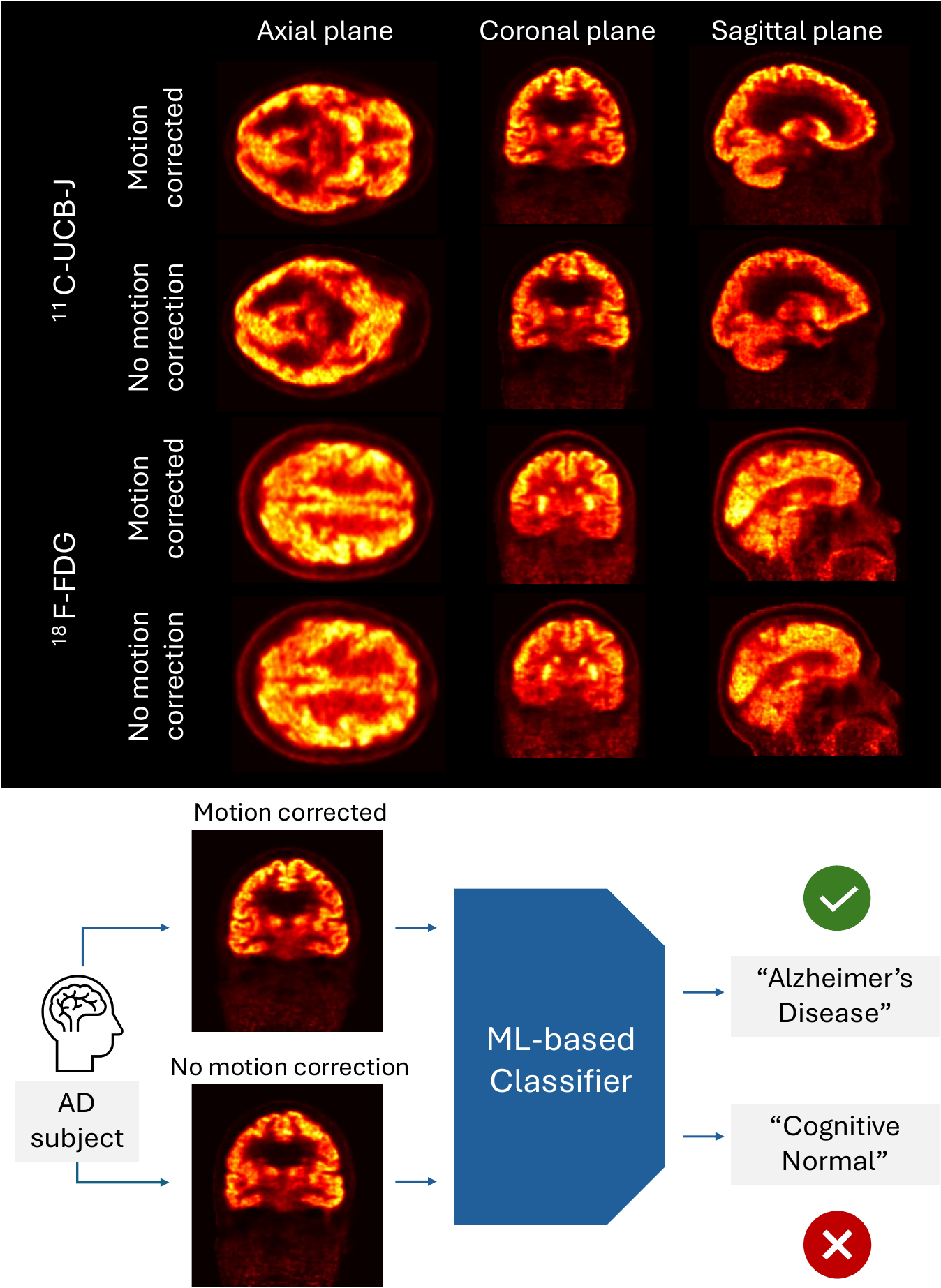}
    \vspace{-2.0\baselineskip}
    \caption{(Top) Comparison of $^{11}$C-UCB-J and $^{18}$F-FDG PET scans reconstructed with and without motion correction in patients with AD.
    (Bottom) Example of $^{11}$C-UCB-J AD patient misclassified as cognitive normal in the absence of motion correction.}
    \label{fig:MC_NMC_recon}
     \vspace{-1.0\baselineskip}
\end{figure}

\end{abstract}
\begin{keywords}
Alzheimer's disease, Brain, PET, Motion Correction, Classification, Supervised Learning, ResNet, SVM, $^{11}$C-UCB-J, $^{18}$F-FDG
\end{keywords}
\section{Introduction}
\label{sec:intro}
Alzheimer's disease (AD) is a neurodegenerative condition characterized by progressive loss of memory and thinking abilities.
AD and related dementia affect 50 million people worldwide and their prevalence is expected to reach 152 million individuals by 2050 in the absence of effective prevention or treatment \cite{Passeri2022Alzheimer}, which makes it a major public health concern.
Brain positron emission tomography (PET) imaging plays a pivotal role in AD diagnosis and staging in clinical routines.
It has the potential to allow for early detection of AD, due to its ability to highlight disease features up to decades before the onset of clinical symptoms \cite{Mosconi2006Hypometabolism}.
In addition, it has a crucial role in deepening our understanding of AD through imaging-based biomarkers and supporting the development of treatment.
In recent years, machine learning (ML) based on non-deep learning, e.g. support vector machine (SVM) of region of interest (ROI) uptake~\cite{Gray2012-qk}, and deep learning~\cite{Liu2018-zx, Singh2017-uj} techniques have been applied to classify cognitive normal (CN), mild cognitive impairment (MCI), and different AD stages from PET studies.
These classification algorithms typically involve collection of multimodal data \cite{Park2023Deep} comprising PET images (tau, amyloid or FDG-PET), neuropsychological evaluation scores (e.g. MMSE), and demographic information that can be part of the publicly-available Alzheimer's Disease Neuroimaging Initiative (ADNI) resources.
They commonly involve cumbersome image preprocessing to compute tracer standardized uptake value ratio (SUVR) for specific brain regions of interest (ROIs), which requires the co-registration of magnetic resonance imaging (MRI) with PET to segment the brain into these ROIs.
Many of these methods have achieved high classification accuracy scores \cite{Park2023Deep} \cite{Azmi2017FDG} \cite{Yang2020Classification}.  
However, as is the case in the majority of brain PET acquired in clinical routines, the ADNI images involved in these studies have limited or no head motion correction applied during image formation.
Motion artifacts significantly degrade image quality (\cref{fig:MC_NMC_recon}) and bias tracer uptake quantification, which can adversely affect the accuracy of disease diagnosis.
In fact, just a 5-mm motion can bias the regional intensity to 35\% \cite{Jin2013evaluation}, which is larger than some differences in regional uptake that distinguish different patient groups. 
We hypothesize that the performance of a ML AD/CN classifier will degrade when used to classify images that are not corrected for patient motion, compared to motion-corrected (MC) images.
To test this hypothesis, we propose training a binary classifier (AD/CN) solely based on PET reconstructions and performing classification on testing data with and without motion correction. 
While the goal of the study was not to achieve state-of-the-art AD classification performance, we nonetheless propose a new ML classification strategy for AD brain PET.
Our results confirmed this hypothesis on 40 testing subjects from two PET tracers used to study AD.
This paper demonstrates the critical importance of correcting motion for PET images to make the most of AD image analysis within the context of ML.
\section{Methods}
\label{sec:format}
\subsection{PET Image Reconstruction}
\label{sec:img}
To reconstruct PET images, we considered list-mode data acquired 40-60 minutes post-injection.
From this data, we reconstructed standardized uptake value (SUV) PET images using motion-compensation OSEM list-mode algorithm for resolution-recovery reconstruction (MOLAR)~\cite{Carson2003Design} (2 iterations $\times$ 30 subsets). 
Each PET data was reconstructed twice, one with motion correction (MC) using Vicra data, and one with no motion correction (NMC).
Normalization, attenuation correction, random correction, scatter correction, and spatially invariant point-spread-function (PSF) of 2.5-mm full-width-half-maximum (FWHM) were applied for both reconstructions.
Before feature extraction, the reconstructions were downsampled from a dimension of 256$\times$256$\times$207 to 
128$\times$128$\times$128 with a final voxel size of 
2.44$\times$2.44$\times$1.98 mm$^3$. 
No post-processing smoothing filter was applied.
\subsection{AD Classification Algorithm}
\label{sec:methods}
Our classification method employs a two-step algorithm that first extracts image features and then performs binary classification into cognitive normal (CN) or AD patients solely based on the PET image.
Unlike most current studies, no additional data other than the PET reconstruction, such as cognitive test scores, other imaging modalities, or lab tests is needed.
The feature extractor utilizes a ResNet10 \cite{He2016Deep} architecture.
We explored different feature extraction training strategies:
\begin{inparaenum}[(i)]
    \item train from scratch using our data and retrieving features extracted by the trained network; 
    \item extract features using ResNet10 pre-trained on MedicalNet~\cite{medicalnet} without any further training (zero-shot learning); and 
    \item fine-tune ResNet10 pre-trained on MedicalNet with our data (multi-shot learning).
\end{inparaenum}
In terms of the classification, we consider the following routines:
\begin{inparaenum}[(i)]
    \item using ResNet10 directly as a binary classifier based on the features extracted; and 
    \item fitting a Support Vector Machine (SVM) classifier on features extracted by ResNet10 on images from training and validation data and then using the SVM classifier for the prediction on testing data.
\end{inparaenum}
\section{Experiments and Results}
\subsection{Experimental Setup}
\paragraph*{Dataset}
From our institution, we identified PET data acquired with two radiotracers that are especially relevant to the study and diagnosis of AD.
We include the following studies:
\begin{inparaenum}[(i)]
    \item 128 $^{11}$C-UCB-J PET scans reflecting synaptic density, the loss of which being a hallmark of AD \cite{Mecca2022synaptic}; and
    \item 173 $^{18}$F-FDG PET scans, which highlight changes in brain glucose metabolism that typically happen in AD \cite{Chtelat2020amyloid}.
\end{inparaenum}
All data was acquired at the Yale PET Center on a brain-dedicated High-Resolution Research Tomograph (Siemens, Germany) and includes both CN and AD patients (\cref{dataset}).
For each PET study, head motion data was recorded using a Polaris Vicra (Northern Digital Inc., Canada) marker-based motion tracking device.
A summary of motion magnitude in the testing cohorts was computed over the whole scan duration~\cite{Jin2013evaluation} 
and was found to be 5.58$\pm$3.89 mm and 7.01$\pm$4.63 mm across the $^{11}$C-UCB-J and $^{18}$F-FDG  testing sets, respectively. 
\begin{table}[t]
\caption{PET Data Cohort.}
\label{dataset}
\resizebox{\columnwidth}{!}{%
\begin{tabular}{lcccccc}
\midrule
                    & \multicolumn{3}{c}{$^{11}$C-UCB-J} & \multicolumn{3}{c}{$^{18}$F-FDG} \\
                    \cmidrule(l){2-4} 
                    \cmidrule(l){5-7} 
                    & Train  & Val. &Test  & Train  & Val. & Test       \\ \midrule
Cognitive Normal    & 47           & 2  & 10            & 94         &  10   & 10             \\
Alzheimer's Disease & 52            & 7  & 10            & 42         &  7  & 10             \\ \midrule
\textbf{Total}      & \textbf{99}  & \textbf{9} & \textbf{20}   & \textbf{136} &\textbf{17}    & \textbf{20}    \\ \midrule
\end{tabular}%
}
\end{table}
\paragraph*{Implementation Details}
To determine the best classification model, we trained the ResNet10 model variants and examined their performance with our two classification strategies on both $^{11}$C-UCB-J and $^{18}$F-FDG datasets. 
We train ResNet10 models (14.4M parameters) for binary classification using 
binary cross-entropy loss, a batch size of 8 for training and validation sets, and Adam optimization using an initial learning rate of 0.001.
In addition to using ResNet directly for binary classification, we also use the model to extract features at different depths and fit an SVM classifier based on training and validation sets features.
Once the features have been extracted from the PET images with ResNet, they are reduced using maximum pooling of the feature maps and then fed into SVM.
We compare SVM performance on features at different ResNet depths.
All models were implemented using Python, PyTorch (v2.4.0) and MONAI (v1.3.2), and trained on an NVIDIA Quadro RTX 8000 GPU.
All code will be made publicly available on GitHub.
\paragraph*{Evaluation Metrics}
We evaluate classification performance by computing accuracy, precision, recall, F1-score, and area under the receiver operating characteristic (AUROC) curve.
Due to limited sample sizes of the test sets, statistical significance testing is not appropriate.
\subsection{Results}
\paragraph*{AD Classification}
We started by studying the effect of the choice of classifier on the classification performance in both cohorts. 
To do so, we extracted features from the training and validation set images (\cref{dataset}) using ResNet10 pre-trained on MedicalNet and obtained features of sizes 
64$\times$32$\times$32$\times$32, 128$\times$16$\times$16$\times$16, 
256$\times$8$\times$8$\times$8 and
512$\times$4$\times$4$\times$4. 
These features computed at different depths in the image encoding process were then used to fit different SVM classifiers.
The same feature extraction process at different depths was performed on testing subjects with and without motion correction, and we used the fitted classifiers to predict if these testing subjects were cognitive normal or AD patients.
We also performed classification based on the full pre-trained ResNet10 model.
We observe that overall, SVM outperforms ResNet at the classification task (\cref{features_ablation}), with ResNet classification accuracy being 50\% on both testing cohorts. 
If we focus on SVM classification performance at different network depths, we observe a decrease in accuracy and other metrics as the number of features increases, likely caused by overfitting.
We noticed that using 128 16$\times$16$\times$16 features leads to the best results on $^{11}$C-UCB-J subjects, with an accuracy of 90\% and identification of all AD cases (Recall = 1). 
On $^{18}$F-FDG subjects, using 64 32$\times$32$\times$32 features yielded the best results, with an accuracy of 85\%. 
We also observe that for both tracers, the overall performance of the classifiers decreases when images are not corrected for motion, except for $^{18}$F-FDG prediction using 256 features.
\begin{table*}[h!]
\caption{Ablation studies on classification method (ResNet10 or SVM fit on different features at different depths) for FDG and UCB-J testing cohorts. The feature extraction method is fixed at ResNet10 pre-trained on MedicalNet without any further training (zero-shot learning).}
\centering
\resizebox{1.0\linewidth}{!}
{
\begin{tabular}{llcc|cc|cc|cc|cc}
\toprule
\multicolumn{1}{l}{\multirow{2}{*}{\textbf{Tracer}}} & 
\multicolumn{1}{l}{\multirow{2}{*}{\textbf{Classification}}} & 
\multicolumn{2}{c|}{\textbf{Accuracy}} & 
\multicolumn{2}{c|}{\textbf{Precision}} & 
\multicolumn{2}{c|}{\textbf{Recall}} & 
\multicolumn{2}{c|}{\textbf{F1-score}} & 
\multicolumn{2}{c}{\textbf{AUROC}} \\ 
\cmidrule(l){3-12} 
\multicolumn{1}{c}{} & \multicolumn{1}{c}{} & ~MC~ & ~NMC~ & ~MC~ & ~NMC~ & ~MC~ & ~NMC~& ~MC~ & ~NMC~ & ~MC~ & ~NMC \\ \midrule
\multirow{5}{*}{\textbf{$^{11}$C-UCB-J}} & ResNet10 & 0.50 & 0.50 & 0.50 & 0.50 & 1.00 & 1.00 & 0.67 & 0.67 & 0.44 & 0.29 \\
& ResNet10 (64 feat.) + SVM & 0.85 & 0.85 & 0.82 & 0.77 & 0.90 & 1.00 & 0.86 & 0.87 & 0.92 & 0.92 \\
& ResNet10 (128 feat.) + SVM & \textbf{0.90} & 0.85 & 0.83 & \textbf{0.89} & \textbf{1.00} & 0.80 & \textbf{0.91} & 0.84 & \textbf{0.96} & 0.91 \\
& ResNet10 (256 feat.) + SVM & 0.70 & 0.65 & 0.67 & 0.62 & 0.80 & 0.8 & 0.73 & 0.70 & 0.77 & 0.82\\
& ResNet10 (512 feat.) + SVM & 0.45 & 0.45 & 0.45 & 0.44 & 0.50 & 0.40 & 0.48 & 0.42 & 0.54 & 0.51 \\\midrule \midrule
\multirow{5}{*}{\textbf{$^{18}$F-FDG}} & ResNet10 & 0.50 & 0.50 & 0.00 & 0.00 & 0.00 & 0.00 & 0.00 & 0.00 & 0.66 & 0.69 \\
& ResNet10 (64 feat.) + SVM & \textbf{0.85} & 0.75 & \textbf{1.00} & 0.86 & 0.70 & 0.60 & \textbf{0.82} & 0.71 & \textbf{0.94} & 0.84 \\
& ResNet10 (128 feat.) + SVM & 0.80 & 0.76 & 0.88 & 0.78 & 0.70 & 0.70 & 0.78 & 0.74 & 0.93 & 0.81 \\
& ResNet10 (256 feat.) + SVM & 0.60 & 0.71 & 0.60 & 0.67 & 0.60 & \textbf{0.80} & 0.60 & 0.73 & 0.75 & 0.83 \\
& ResNet10 (512 feat.) + SVM & 0.65 & 0.62 & 0.71 &  0.62 & 0.50 & 0.50 & 0.59 & 0.56 & 0.70 & 0.59 \\
\bottomrule
\end{tabular}
}
\label{features_ablation}
\end{table*}
\paragraph*{Feature Extraction}
We then quantified classifier performance using different methods for feature extraction, namely training ResNet10 for classification from scratch on our data, using pre-trained ResNet10 without further training, and fine-tuning this pre-trained model to our data (\cref{feature_extraction_ablation}). 
We chose a baseline classifier based on what we observed to be the best option in our first ablation study, which was to use 128 features and SVM for $^{11}$C-UCB-J, and 64 features with SVM on $^{18}$F-FDG. 
\begin{table*}[h!]
\centering
\caption{Feature extractor ablation studies results on the UCB-J and FDG testing cohorts. For SVM fitting, we consider 128 and 64 features for UCB-J and FDG, respectively and try various feature extraction strategies.}
\label{feature_extraction_ablation}
\resizebox{1.0\linewidth}{!}
{
\begin{tabular}{llcc|cc|cc|cc|cc}
\toprule
\multicolumn{1}{l}{\multirow{2}{*}{\textbf{Tracer}}} & 
\multicolumn{1}{l}{\multirow{2}{*}{\textbf{Feature extractor}}} & 
\multicolumn{2}{c|}{\textbf{Accuracy}} & 
\multicolumn{2}{c|}{\textbf{Precision}} & 
\multicolumn{2}{c|}{\textbf{Recall}} & 
\multicolumn{2}{c|}{\textbf{F1-score}} & 
\multicolumn{2}{c}{\textbf{AUROC}} \\ 
\cmidrule(l){3-12} 
\multicolumn{1}{c}{} & \multicolumn{1}{c}{} & ~MC~ & ~NMC~ & ~MC~ & ~NMC~ & ~MC~ & ~NMC~& ~MC~ & ~NMC~ & ~MC~ & ~NMC \\ \midrule
\multirow{3}{*}{\textbf{$^{11}$C-UCB-J}} & ResNet10 trained on our data & 0.70 & 0.60 & 0.70 & 0.58 & 0.70 & 0.70 & 0.70 & 0.64 & 0.72 & 0.65 \\
& Pre-trained ResNet10 (zero-shot) & \textbf{0.90} & 0.85 & 0.83 & \textbf{0.89} & \textbf{1.00} & 0.80 & \textbf{0.91} & 0.84 & \textbf{0.96} & 0.91 \\
& Pre-trained ResNet10 (fine-tuned) & 0.65 & 0.75 & 0.64 & 0.69 & 0.70 & 0.90 & 0.67 & 0.78 & 0.79 & 0.87 \\ \midrule \midrule
\multirow{3}{*}{\textbf{$^{18}$F-FDG}} & ResNet10 trained on our data & 0.80 & 0.75 & 0.88 & 0.86 & 0.70 & 0.60 & 0.78 & 0.71 & 0.90 & 0.88 \\
& Pre-trained ResNet10 (zero-shot) & \textbf{0.85} & 0.75 & \textbf{1.00} & 0.86 & \textbf{0.70} & 0.60 & \textbf{0.82} & 0.71 & 0.94 & 0.84 \\
& Pre-trained ResNet10 (fine-tuned) & \textbf{0.85} & 0.75 & \textbf{1.00} & 0.78 & \textbf{0.70} & 0.70 & \textbf{0.82} & 0.74 & \textbf{0.96} & 0.88 \\
\bottomrule
\end{tabular}
}
\end{table*}

\begin{figure}[t]
    \centering
    \includegraphics[width=\linewidth]{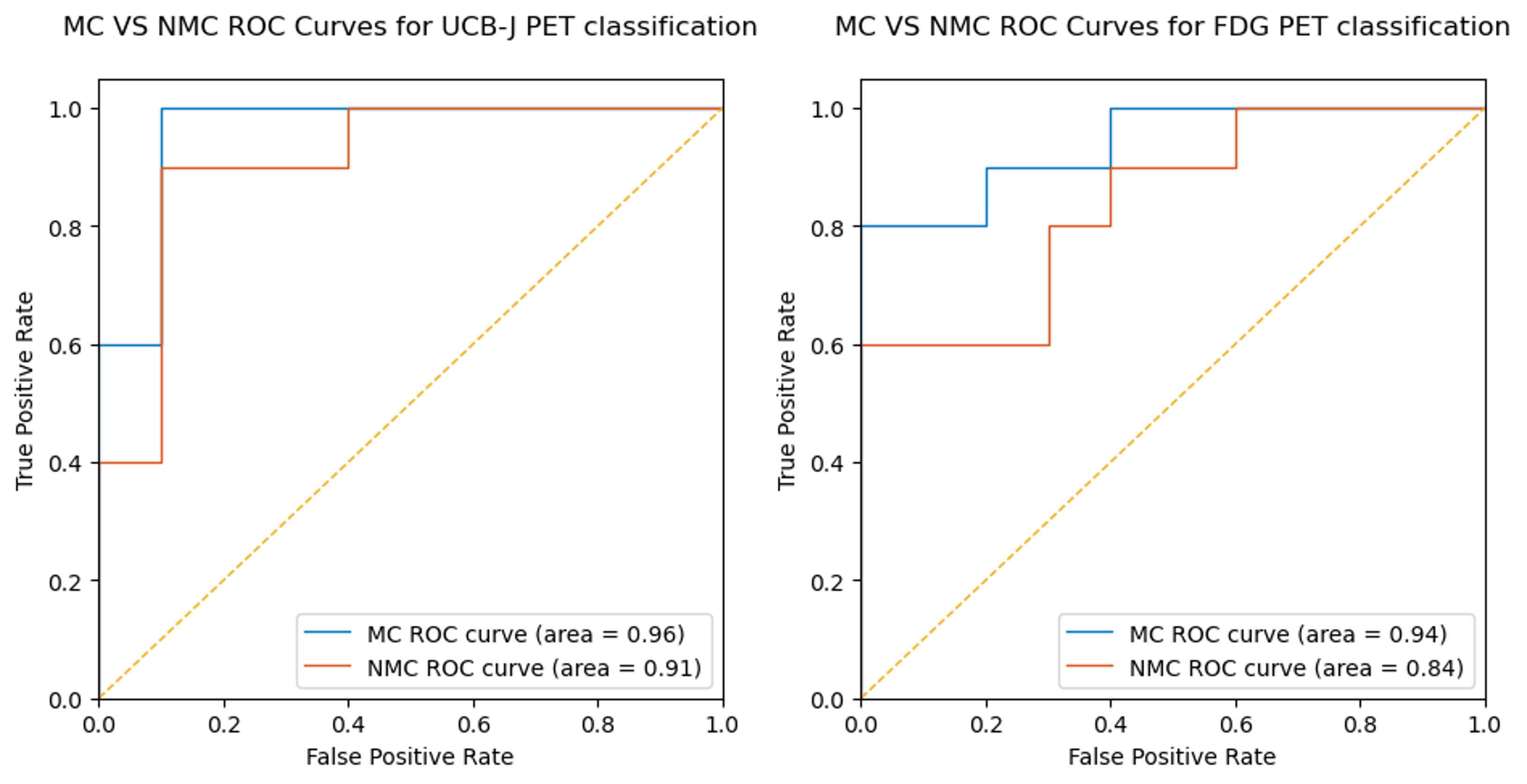}
    \caption{AUROC curves comparing the best model performances on testing subjects with motion correction (MC) and without motion correction (NMC).}
    \label{fig:ROC_curves}
     \vspace{-1.0\baselineskip}
\end{figure}

During model training, we observed that the latter was prone to overfitting to the training data, despite carefully monitoring training with a validation set, which could explain why training on our data does not lead to the best performance.
There is indeed a 20\% difference in accuracy between training ResNet10 on our data and using the pre-trained model in $^{11}$C-UCB-J data, and a 5\% difference in $^{18}$F-FDG subjects.
Overfitting could also explain why fine-tuning the pre-trained ResNet10 fails to outperform the pre-trained model on $^{11}$C-UCB-J data, on which zero-shot learning is the best strategy.
On $^{18}$F-FDG data, results from zero-shot and multi-shot learning are very close with accuracies of 85\% and precision of 100\% for both methods.
\paragraph*{Impact of Head Motion on Classification}
Classifiers reached excellent AUROC scores of 0.96 on FDG and 0.94 on $^{11}$C-UCB-J data on MC images (\cref{fig:ROC_curves}), which is comparable to results found in the literature \cite{Park2023Deep}. 
Our algorithm accurately identifies all AD cases (Recall = 1) on $^{11}$C-UCB-J MC images, while it misses 20\% of them when motion is not corrected.
Recall also drops 10\% when motion is not corrected in $^{18}$F-FDG images.
\Cref{fig:MC_NMC_recon} shows an example of an AD case misclassified as cognitive normal in the absence of motion correction but correctly categorized as AD using the MC reconstruction.
When motion is not corrected, we observe overall classification accuracy drops of 10\% and 5\% in the $^{18}$F-FDG and $^{11}$C-UCB-J data, respectively, for the best performing models. 
In most ablation studies, we observe a decrease in classification-related metrics between MC and NMC images, especially on $^{18}$F-FDG data, which confirms our initial hypothesis.

\section{Discussion and Conclusion}
In this study, we investigated binary classification strategies for brain PET images and demonstrated that their performance is sensitive to motion correction.
We validated our hypothesis using two cohorts of PET images relevant to the study of AD, which total 301 PET scans.
Among ResNet10 training strategies, pre-training on MedicalNet combined with zero-shot learning was shown to be the most effective. 
We hypothesize that the MedicalNet pre-training on a variety of imaging modalities and target organs allows ResNet to extract meaningful features from our PET images while avoiding overfitting induced by training or fine-tuning on our data.
This network generalizes well to both tracers and when combined with SVM for binary classification, achieves high accuracy when fit to 128 features for $^{11}$C-UCB-J and 64 features for $^{18}$F-FDG data.
We observe that classification is overall less accurate on $^{18}$F-FDG images than $^{11}$C-UCB-J images independently of the presence of motion correction. 
We hypothesize that this is inherent to the tracer's nature, $^{11}$C-{UCB-J} being potentially more specific to AD than $^{18}$F-FDG which highlights global and subtle changes in glucose metabolism.
Our results confirmed our hypothesis that head motion impacts classification performance of AD in both cohorts, as we observed 10\% and 5\% drops in accuracy in 20 $^{18}$F-FDG and 20 $^{11}$C-UCB-J testing subjects, respectively. 
To our knowledge, no study has quantified the impact of head motion on disease detection until now, and this is also the first study to perform AD/CN classification in synaptic density PET. 
As PET scanner resolution continues to improve and the development of tracers that target smaller regions of the brain progresses, we anticipate that the impact of motion could become even more pronounced. This is particularly relevant for the early detection of tau protein deposition, a critical aspect in the diagnosis of AD. 
These results stress the need for portable and reliable motion correction methods in brain PET imaging to fully take advantage of machine learning-based diagnosis algorithms.
In future work, we will improve the accuracy of our image-based classification method and expand it to include other forms of dementia such as mild cognitive impairment (MCI), and intermediate AD stages. 
We will examine again how the absence of motion correction affects classification performance and differential diagnosis of these  neurodegenerative diseases.
\section{Compliance with ethical standards}
This HIPAA-compliant retrospective, single-institution study was conducted in accordance with the Declaration of Helsinki, and approval was granted by the Institutional Review Board
of the Yale University School of Medicine with waiver of informed consent.
\section{Acknowledgments}
No funding was received to conduct this study. The authors have no relevant financial or non-financial interests to disclose.
\bibliographystyle{IEEEbib}
\bibliography{References}
\end{document}